\documentclass{aastex}          

\usepackage{epsfig}
\usepackage{graphicx}
\usepackage{color}

\begin{document}
%
\title{ Seismology of solar spicules based on Hinode/SOT observations}

\shorttitle{Seismology of solar spicules}
\shortauthors{Abbasvand et al.}

\author{V.~Abbasvand, H.~Ebadi and Z.~Fazel}
\affil{Astrophysics Department, Physics Faculty,
University of Tabriz, Tabriz, Iran\\
e-mail: \textcolor{blue}{hosseinebadi@tabrizu.ac.ir}}

\begin{abstract}
We analyze the time series of \mbox{Ca\,\textsc{ii}} H-line obtained from \emph{Hinode}/SOT on the solar limb.
The time-distance analysis shows that the axis of spicule undergos quasi-periodic transverse displacement.
We determined the period of transverse displacement as \textbf{$\sim$} $40-150$ s
and the mean amplitude as \textbf{$\sim$} $0.1-0.5$ arc\,sec. For the oscillation wavelength of $\lambda$
\textbf{$\sim$} 1/0.06 arc\,sec \textbf{$\sim$} 11500 km, the estimated kink speed is ${\sim}$ $13$--$83$ km s$^{-1}$.
We obtained the magnetic field strength in spicules as $B_{0}$ = $2$ -- $12.5$ G and the energy flux as $7$ -- $227$ J\,m$^{-2}$\,s$^{-1}$.

\end{abstract}


\section{Introduction}
\label{sec:intro}
Spicules can be seen almost everywhere at the solar limb. They are highly-dynamic, thin, jet-like features in the solar chromosphere. Spicules have been observed for a very long time and have been the subject of numerous reviews \citep{bek68, bek1972, Suematsu1995, Sterling2000}. Perhaps the most interesting aspect about spicules is their potential to mediate the transfer of energy and mass from the photosphere to the corona. This potential has been recognized early \citep{bek68, Pneuman78, Athay82}, but the lack of high-quality observations prevented a better understanding of spicules and their link to the corona \citep{Withbroe83}. The advent of the Hinode Solar Optical Telescope(\emph{SOT}) \citep{Kosugi07, Tsuneta08, Suematsu2008} provided a quantum leap in the understanding of spicules and their properties. With its seeing free high spatial and temporal resolution observations of the chromosphere, Hinode showed much more dynamic spicules than previously thought and re-ignited the discussion on their potential to heat the corona \citep{De2007,De2009,De2011}. While there are several recent studies of the properties of spicules using Hinode/SOT \citep{De2007,Anan2010,Zhang2012}, they are either preliminary results papers or cover only one or two data sets.

It is encouraging that the results from the various groups agree quite well, despite the difficulties in obtaining reliable magnetic field measurements at the limb. \citet{Lopez2005} used full Stokes polarimetry and Hanle effect modelling of He I D3 spectra and find that the magnetic field strength is mostly 10 G at 3500 km height, with some spicules possibly showing fields up to 40 G. \citet{Trujillo2005} used both the Hanle and Zeeman effects and full Stokes polarimetry of the He I 10830 line and find 10 G at 2000 km and again, some spicules with field strengths up to 40 G. Doppler velocities of solar limb spicules show oscillations with periods of $20-55$ and $75$ --− $110$ s. There is also clear evidence of 3-min oscillations at the observed heights. The observed upwards steady flows are $20$ --– $25$ km s$^{-1}$ with short periods ($20-55$ s) in type I spicules. Type I spicules have typical lifetimes of $150-400$ s and maximum ascending velocities of $15-40$ km s$^{-1}$. Type II Spicules in the quiet sun have lifetimes between $50-150$ s, and slightly less in coronal holes. They have a maximum velocity between $40-100$ km s$^{-1}$, with a very small amount with velocities greater than $150$ km s$^{-1}$ \citep{bek68}.

All spicule oscillations events are summarized in a recent review by \citet{Zaqarashvili2009}. They suggested that the observed oscillation periods can be formally divided in two groups: those with shorter periods ($<\!\!2$ min) and those with longer periods ($\geqslant \!\!2$ min)\citep{Zaqarashvili2009}. The most frequently observed oscillations lie in the period ranges of $3-7$ min and $50-110$ s. These spiky dynamic jets are propelled upwards at speeds of about $20-25$ km s$^{-1}$ from photosphere into the magnetized low atmosphere of the sun. Their diameter varies from spicule to spicule having the values from 400 km to 1500 km. The mean length of classical spicules varies from 5000 km to 9000 km, and their typical life time is $5-15$ min.

The typical electron density at heights where the spicules are observed is $3.5 \times 10^{-10}$ -- $2 \times 10^{-11}$ kg m$^{-3}$, and their temperatures are estimated $5000-8000$K  \citep{bek68}. Their periods are estimated $20-55$ and $75 - 110$ s. \citet{De2007}, based on Hinode observations concluded that the most expected periods of transverse oscillations lay between $100$ --– $500$ s, which interpreted as signatures of Alfv\'{e}n waves. Phase speed begins from \textbf{$\sim$} $40$ km s$^{-1}$  at lower heights and reaches to the maximum value of \textbf{$\sim$} 90 km s$^{-1}$ at \textbf{$\sim$} $2500-3000$ km. Then, it decreases to the minimum value of \textbf{$\sim$}$20-25$ km s$^{-1}$ at \textbf{$\sim$}$3500-4500$ km \citep{Ebadi2014}. Therefore, the observed quasi-periodic displacement of spicule axis can be caused due to fundamental standing mode of kink waves. The energy flux storied in the oscillation is estimated as 150 J m$^{-2}$ s$^{-1}$, which is of the order of coronal energy losses in quiet Sun regions \citep{Ebadi2014}. Spicule seismology, which means the determination of spicule properties from observed oscillations and was originally suggested by \citep{Zaqarashvili2009}, has been significantly developed during last years \citep{Ebadi2014,Verth2011,Ebadi2014a}.

In the present work, we study the observed oscillations
in the solar spicules through the data obtained from Hinode.
We trace the spicule axis oscillations via time slice diagrams for determining  period, amplitude, phase speed, magnetic field, Alfv\'{e}n speed and energy flux of the studied spicules. Observed waves can be used as a tool for spicule seismology. A description of the observations and image processing is made in Section~\ref{sec:observations} and in Section~\ref{sec:Results} we discuss spicules properties. Finally, our conclusions are
summarized in Section~\ref{sec:concl}.

\section{Observations and data processing}
\label{sec:observations}
We used a time series of \mbox{Ca\,\textsc{ii}} H-line (396.86 nm) obtained on 22 January, 2007, during 23:26 to 23:42 UT by the Solar Optical Telescope onboard \citep{Tsuneta08}. The spatial resolution reaches 0.2 arc\,sec (\textbf{$\sim$}150 km) and the pixel size is 0.109 arc\,sec (\textbf{$\sim$}80 km) in the \mbox{Ca\,\textsc{ii}} H-line. The time series has a cadence of 20 seconds with an exposure time of 0.5 seconds. The position of $X$--$center$ and $Y$--$center$ of slot are 945 arc\,sec and 0 arc\,sec, while X-FOV and Y-FOV are 223 arc\,sec and 112 arc\,sec, respectively.

The ``fgprep'',``fgrigidalign''(available in solarsoft, http://www.lmsal.com/solarsoft) and ``madmax'' IDL
algorithms are used to reduce the images spikes and jitter effect, to align
time series, and to enhance the finest structures, respectively \citep{Shimizu2008,Koutchmy89}.

\begin{figure*}
\centering
\includegraphics[height=3.5cm, width=15cm]{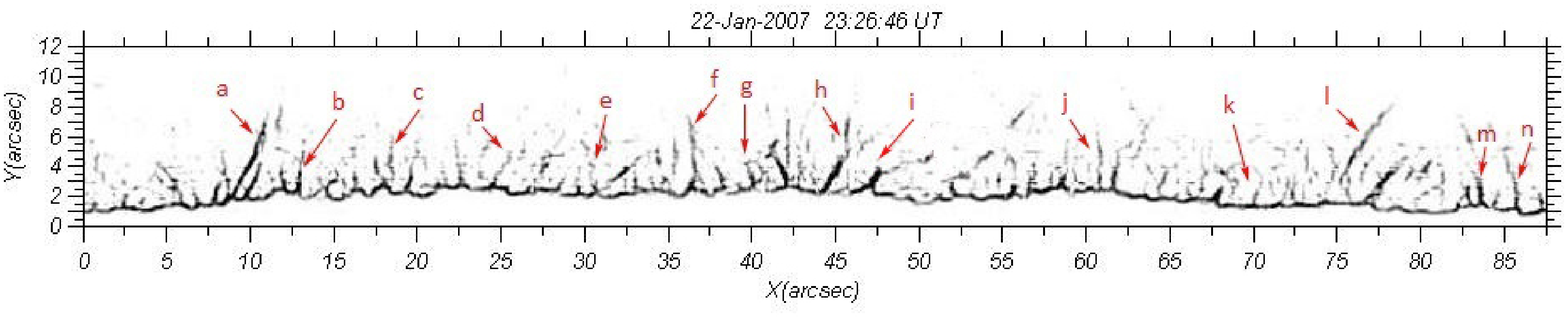}
\caption{The full view \mbox{Ca\,\textsc{ii}} H-line image of the solar equator  which
contains the processed spicules. The red arrows show the studied
spicules. We used the ``madmax'' algorithm
to enhance the finest structures. \label{fig1}}
\end{figure*}
\section{Results and Discussions}
\label{sec:Results}
In Figure~\ref{fig1}, we presented the full view Ca II H-line image of the equator which contains the studied spicules. The processed spicules are indicated by red arrows on this image. This image was taken by Hinode/SOT telescope on 22 Jan, 2007, 23:26:04 UT. We used the time series images of this region and determined period, amplitude, phase speed, magnetic field, Alfv\'{e}n speed and energy flux of the studied spicules. To this end, we selected 14 random spicules and used time-slice diagrams to illustrate their transversal oscillations.

\begin{figure*}
\centering
\includegraphics[height=3cm,width=5cm]{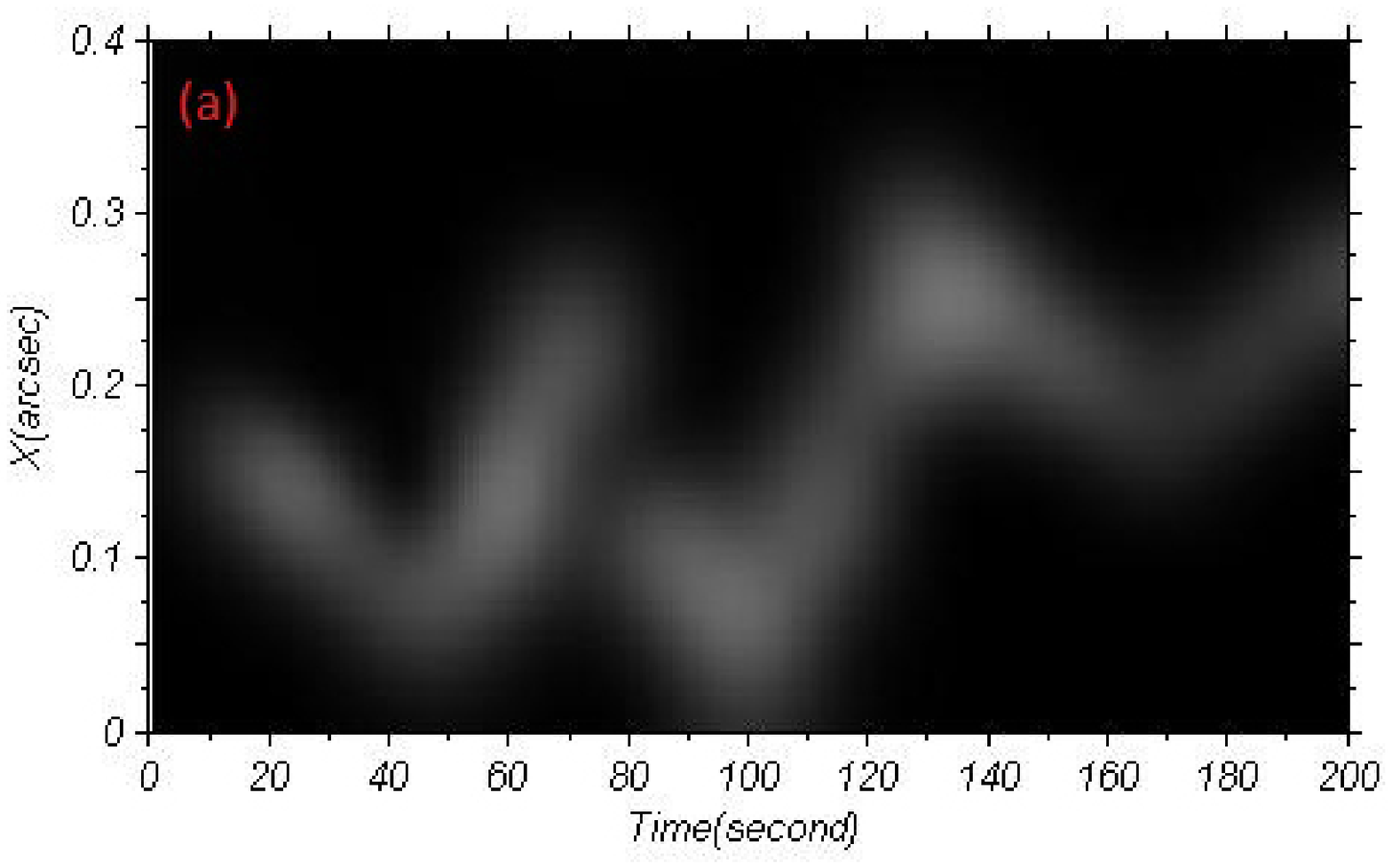}
\includegraphics[height=3cm,width=5cm]{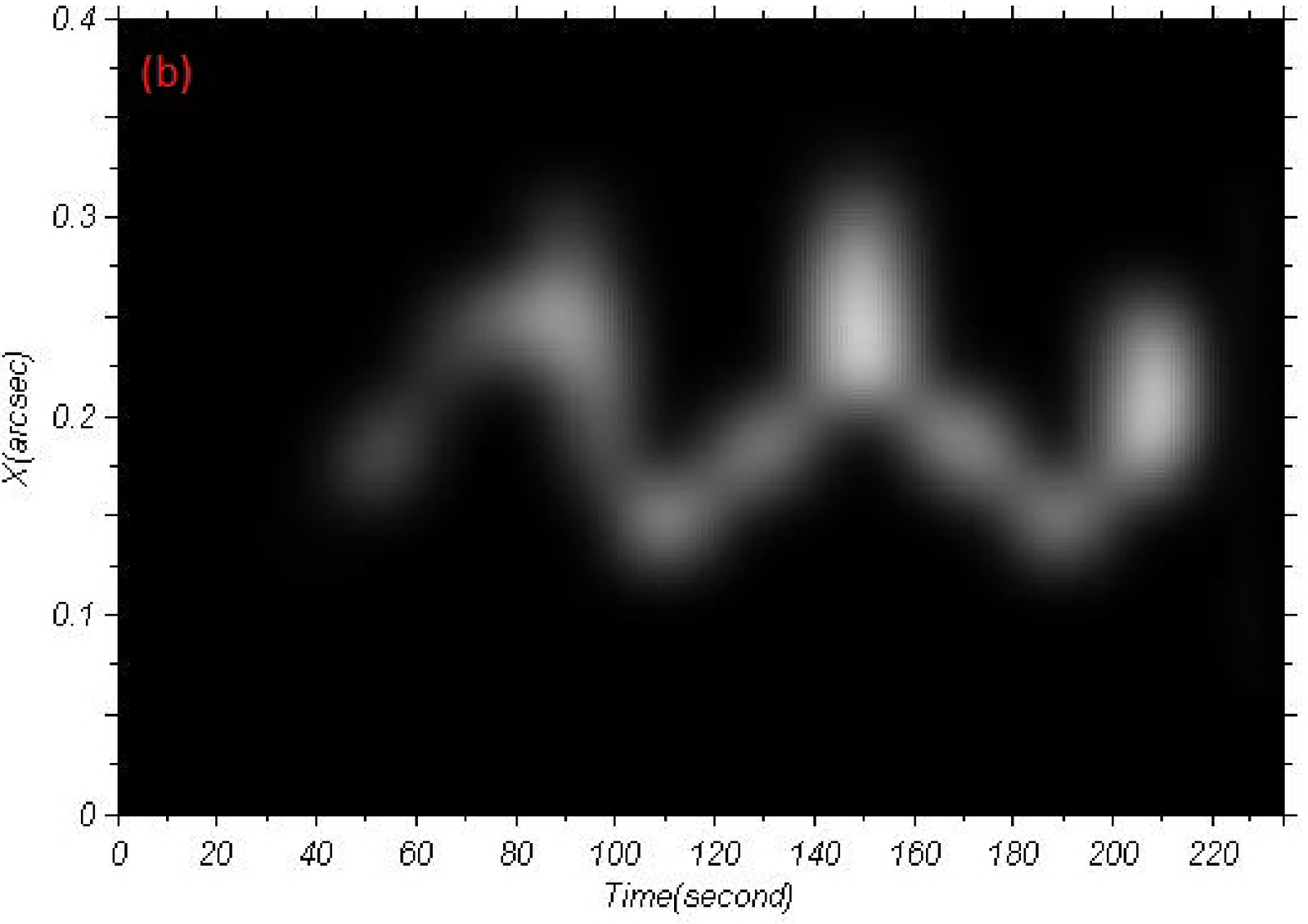}
\includegraphics[height=3cm,width=5cm]{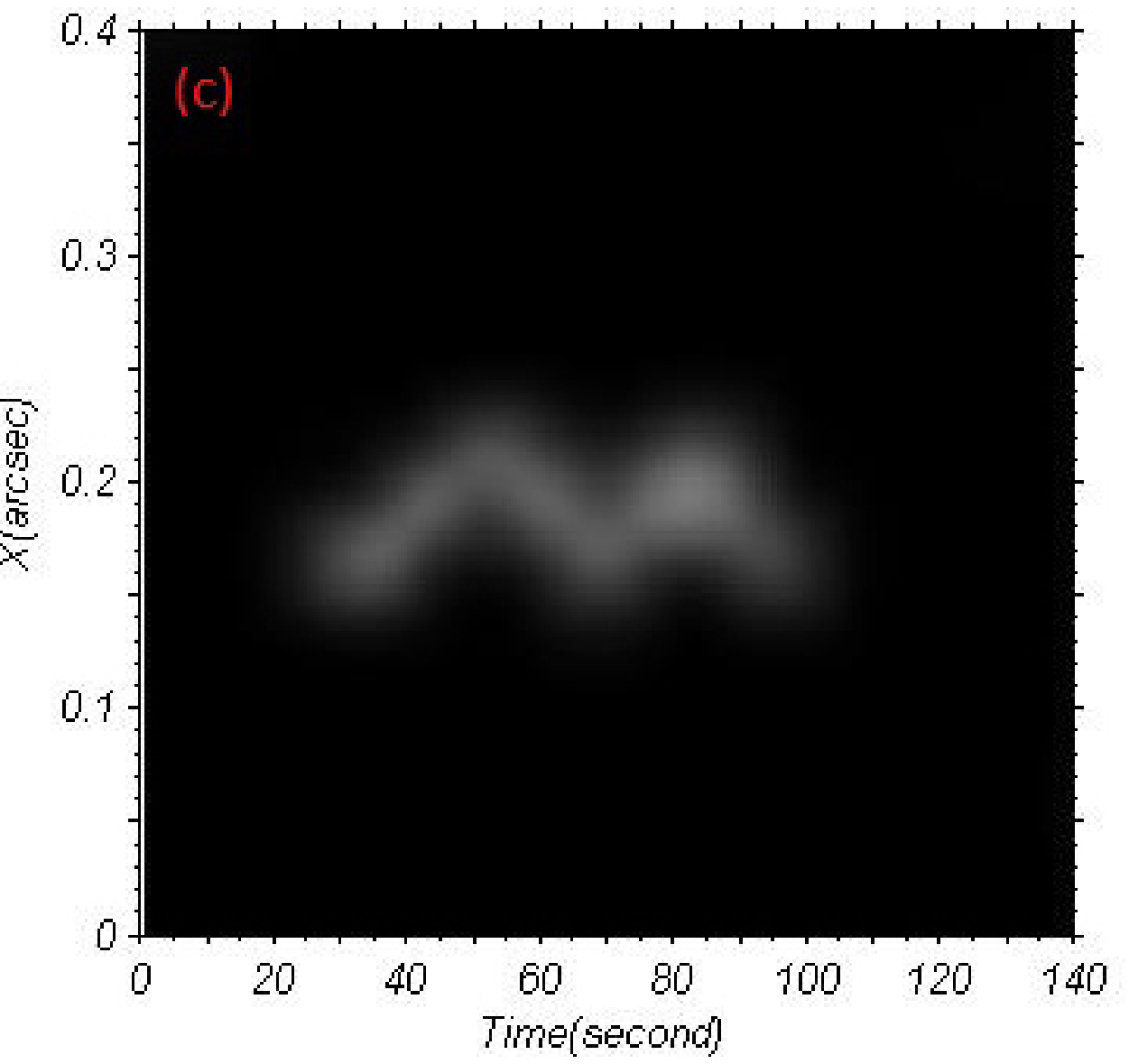}
\includegraphics[height=3cm,width=5cm]{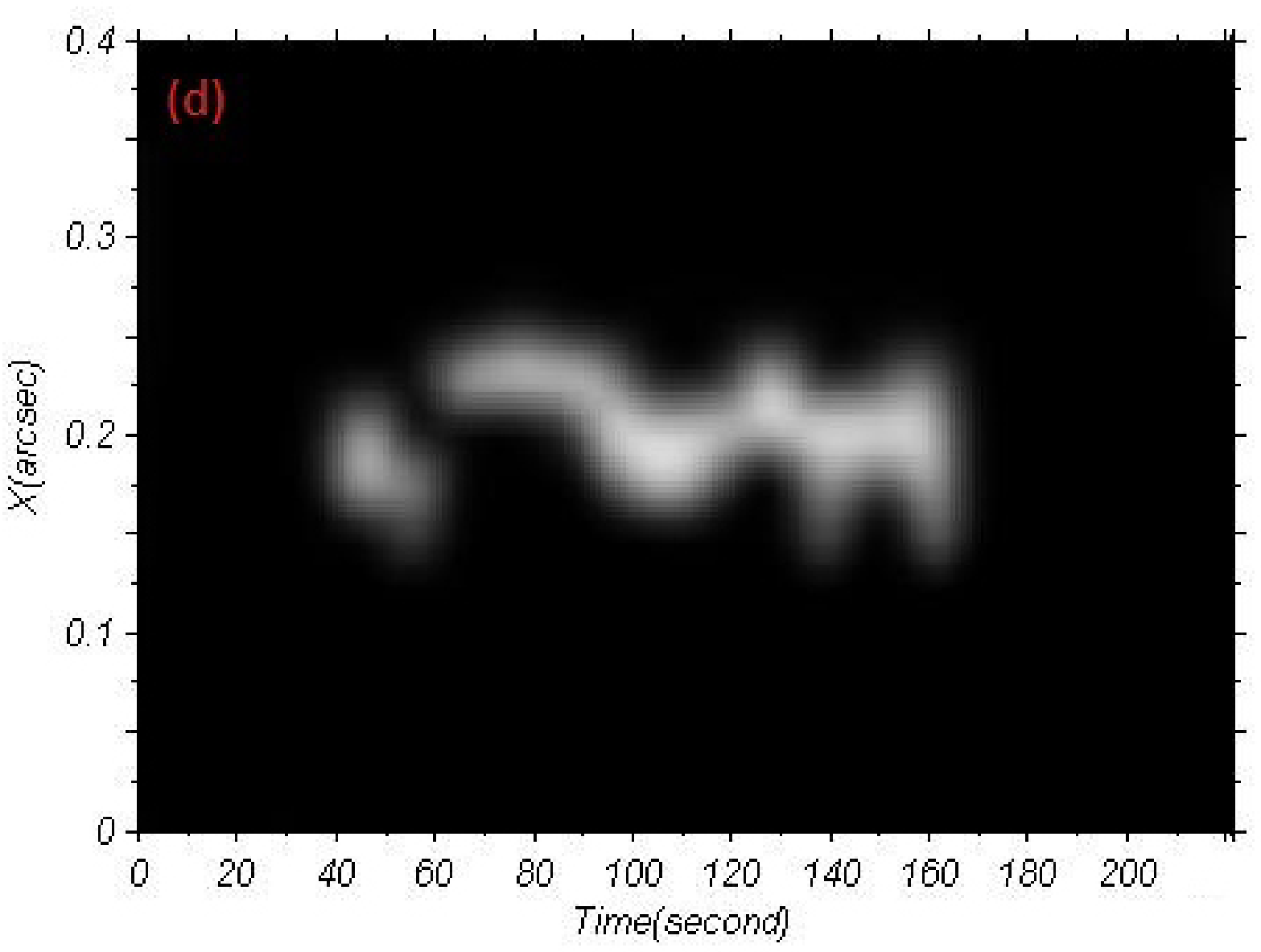}
\includegraphics[height=3cm,width=5cm]{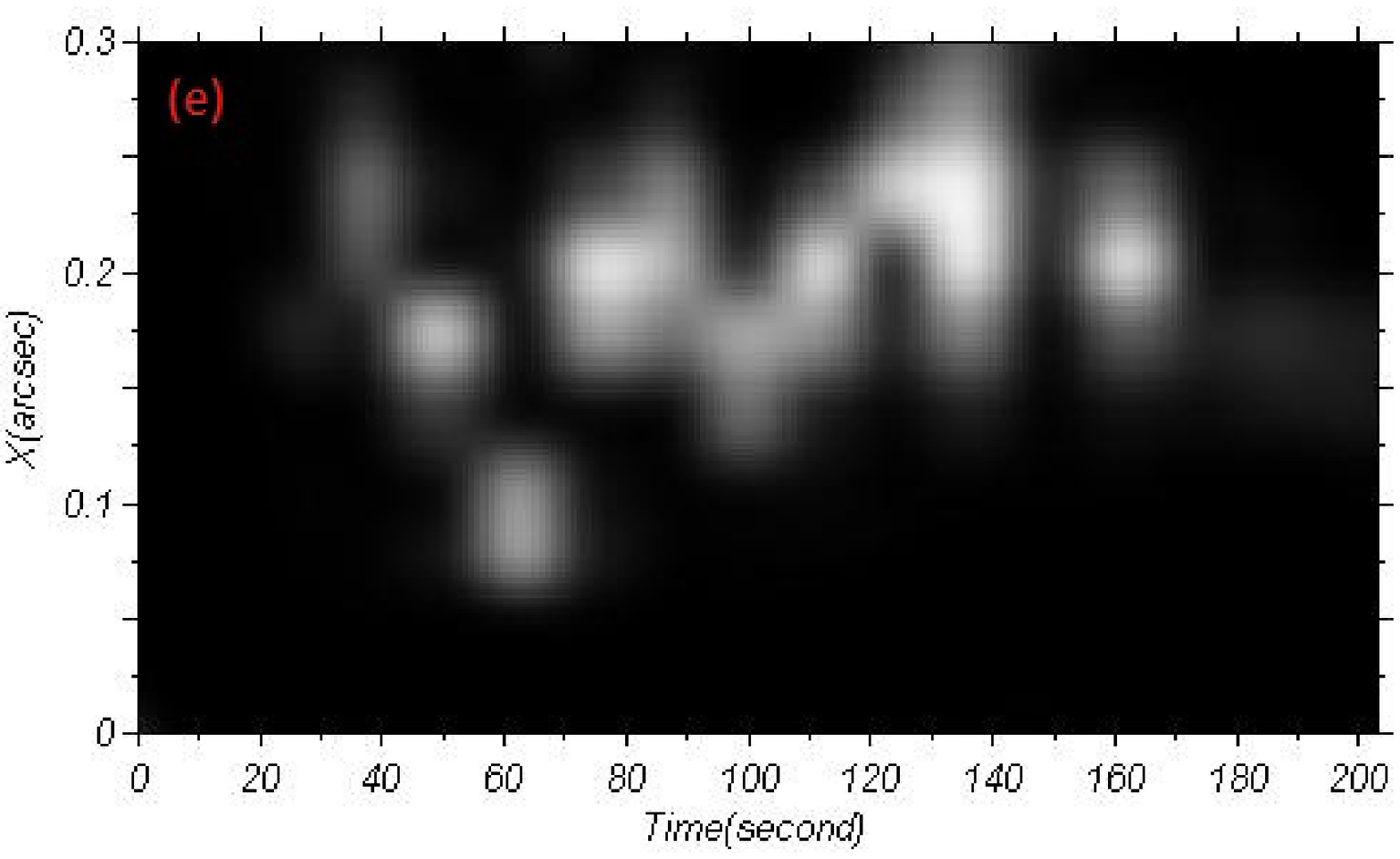}
\includegraphics[height=3cm,width=5cm]{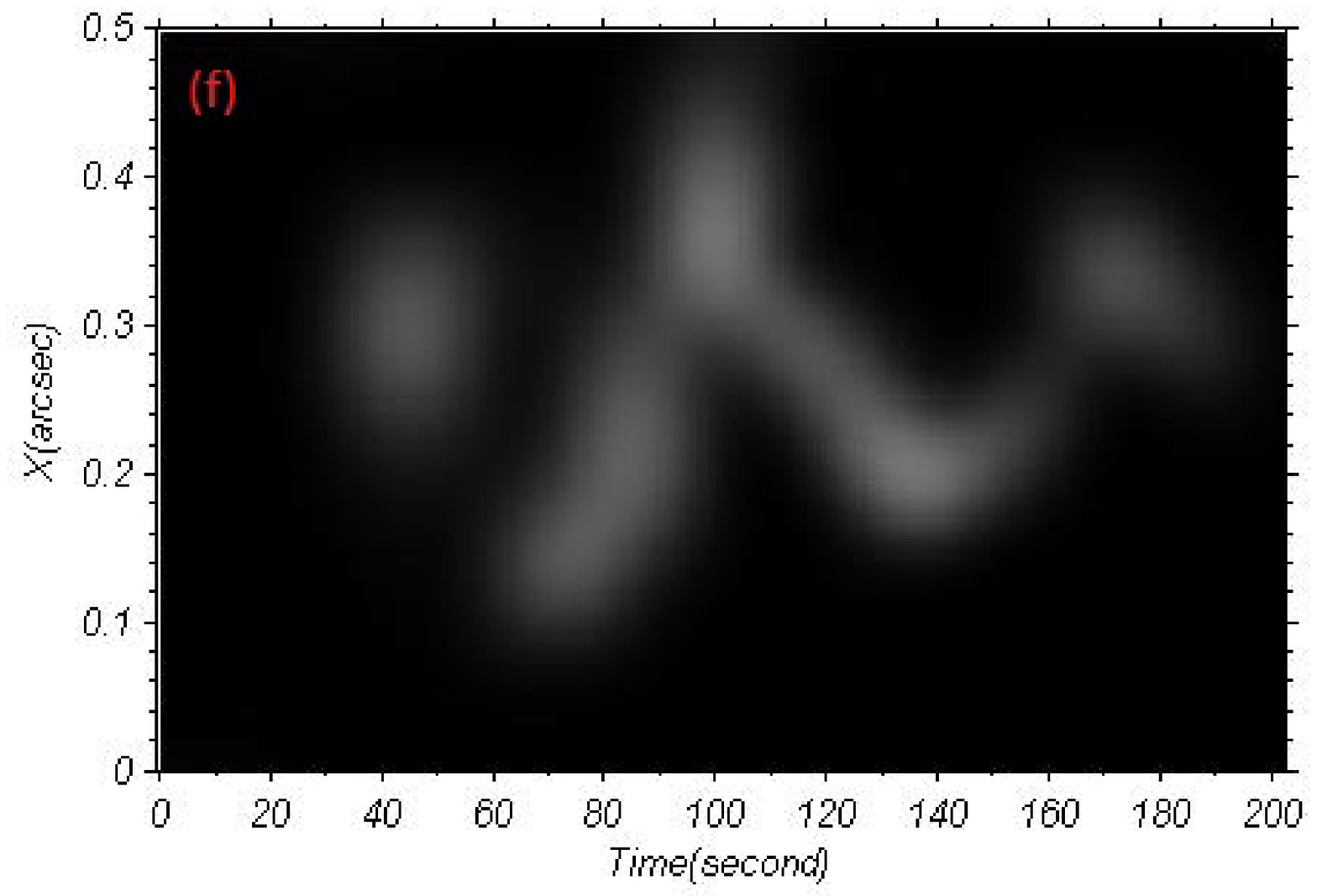}
\includegraphics[height=3cm,width=5cm]{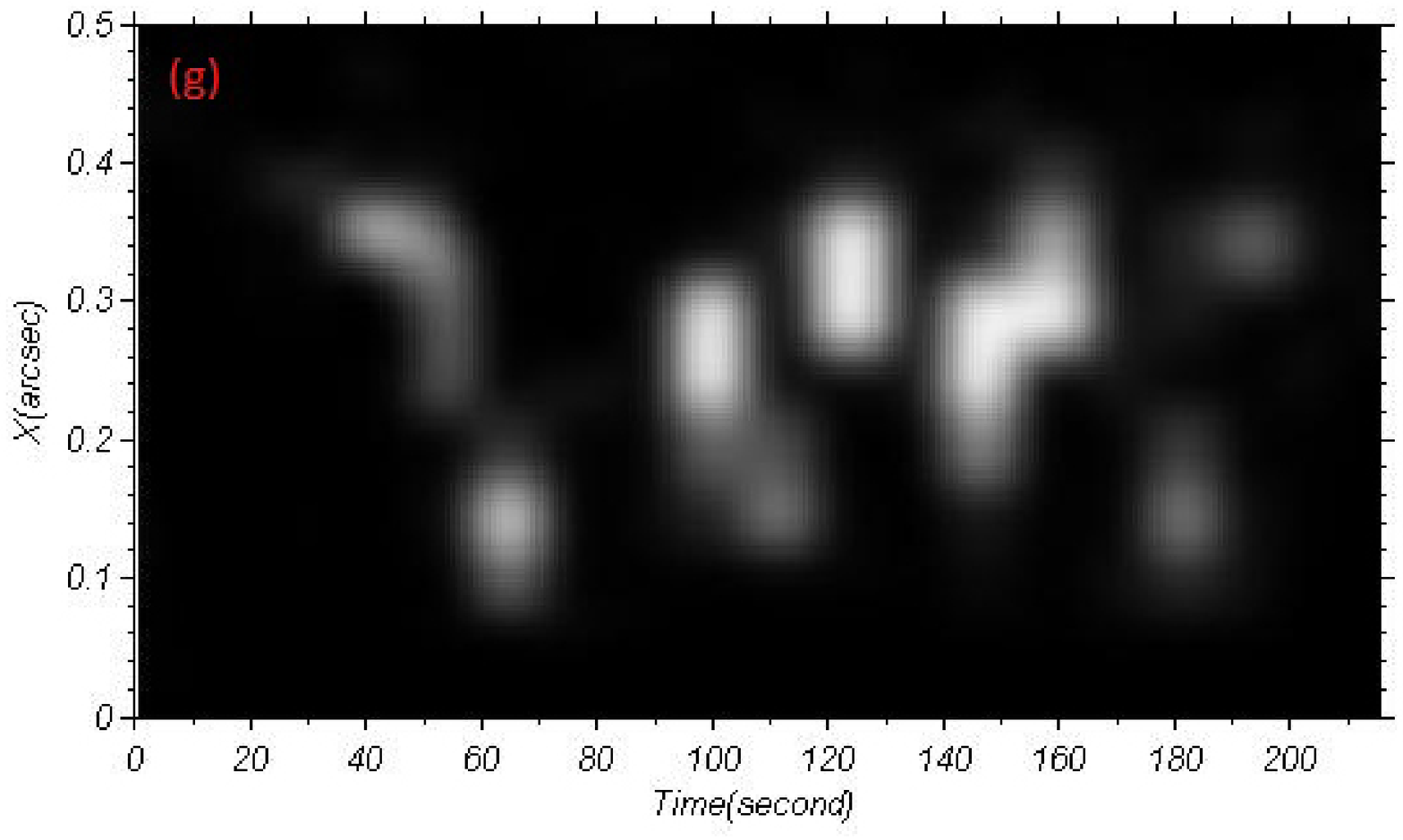}
\includegraphics[height=3cm,width=5cm]{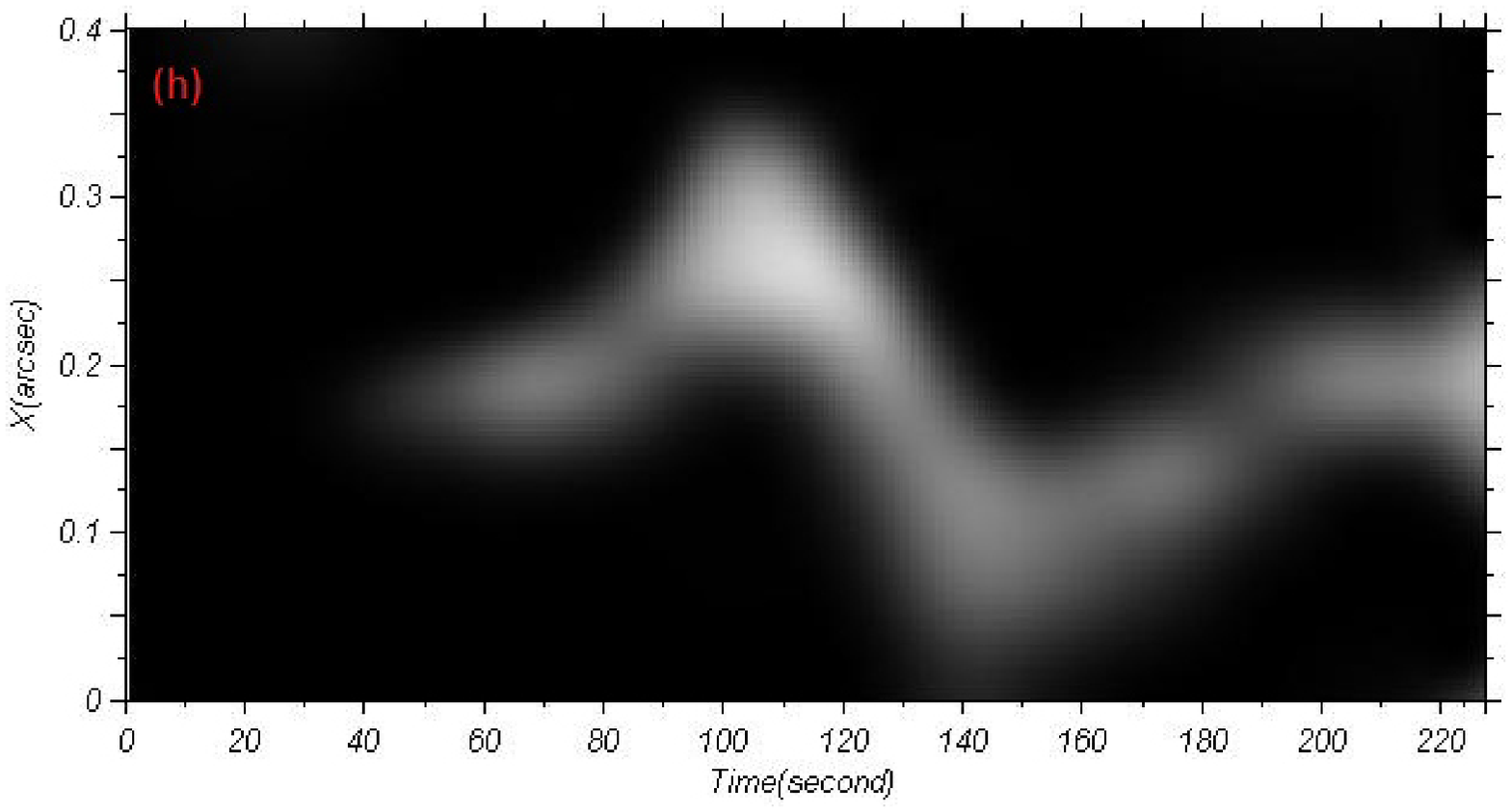}
\includegraphics[height=3cm,width=5cm]{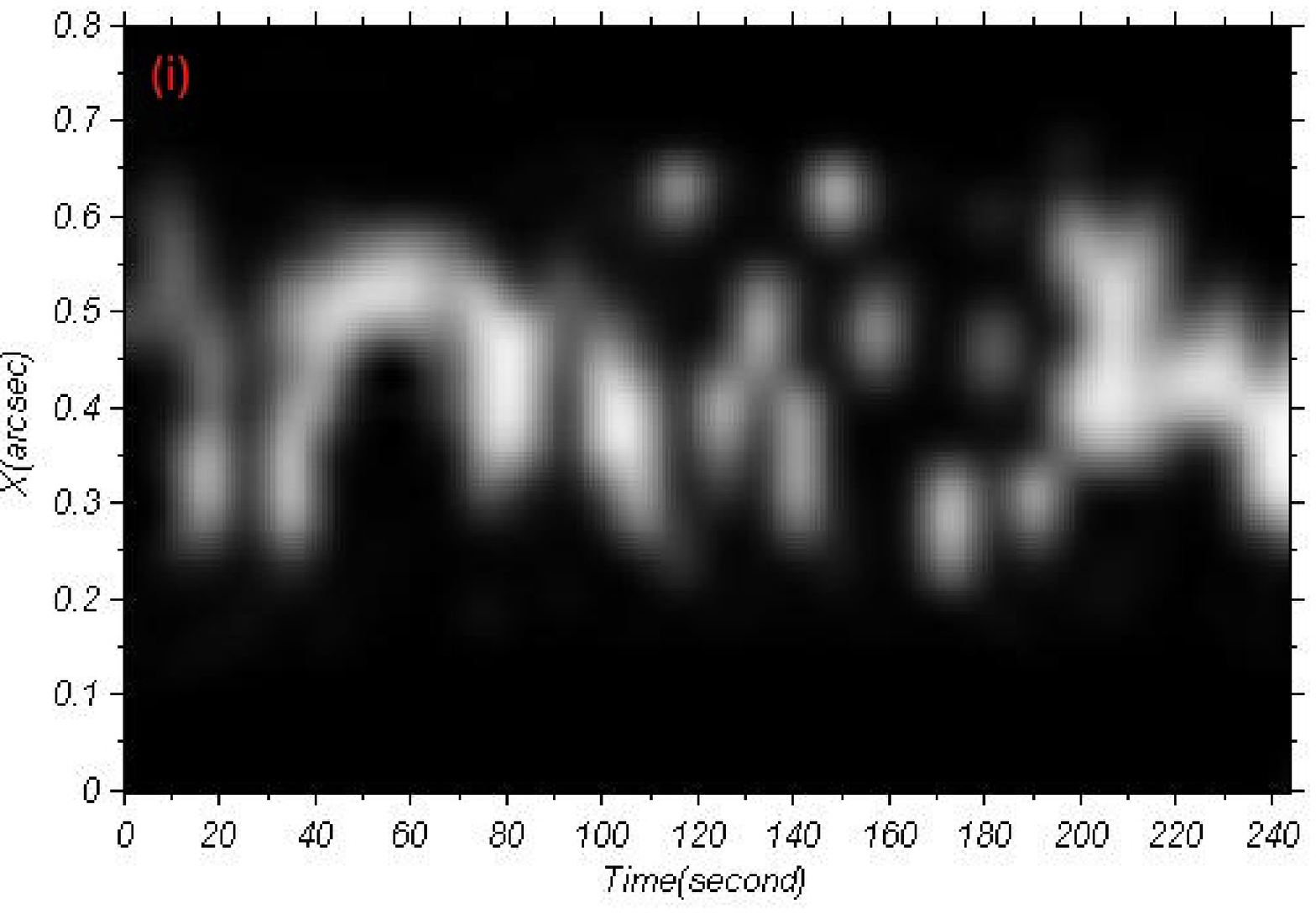}
\includegraphics[height=3cm,width=5cm]{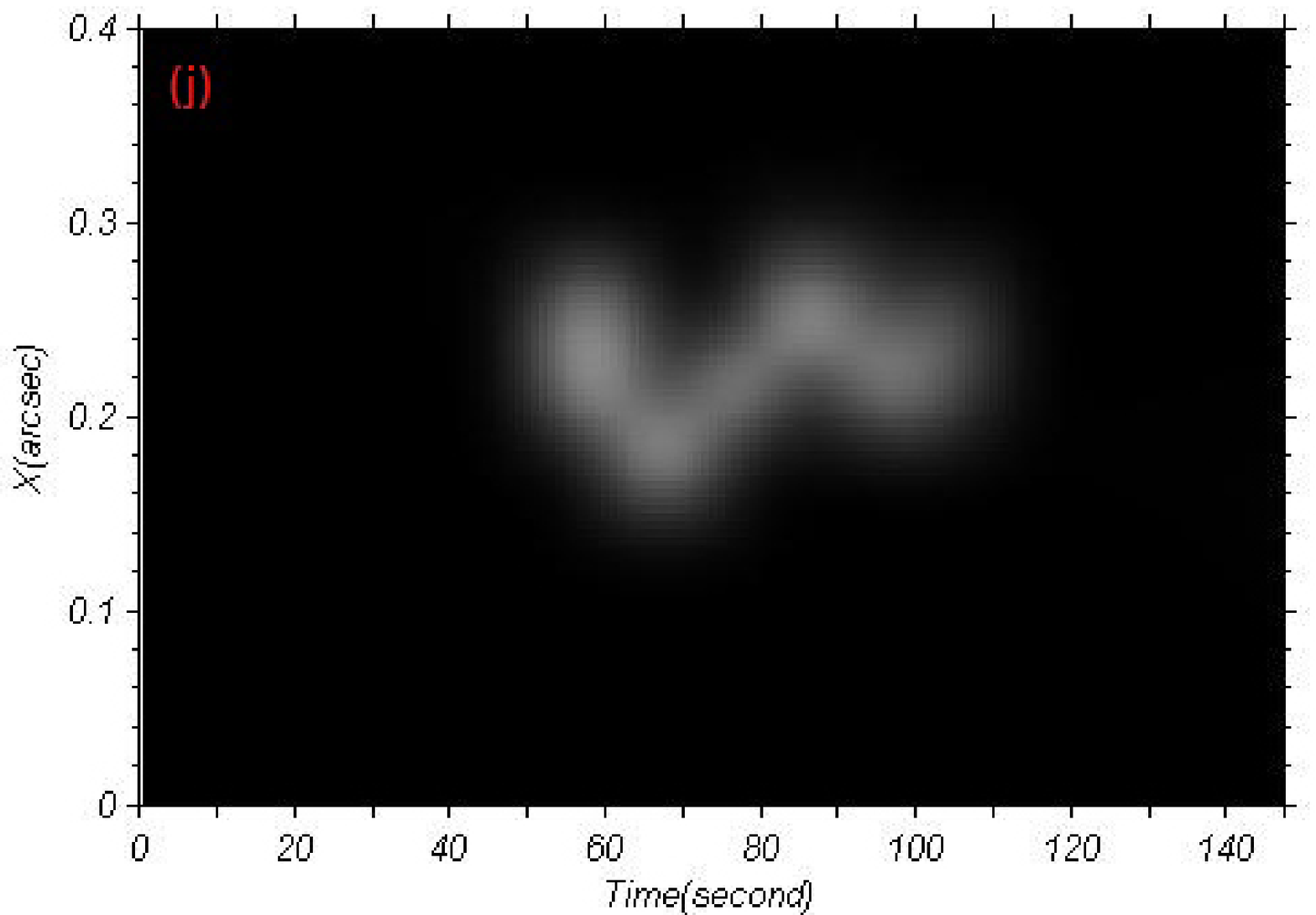}
\includegraphics[height=3cm,width=5cm]{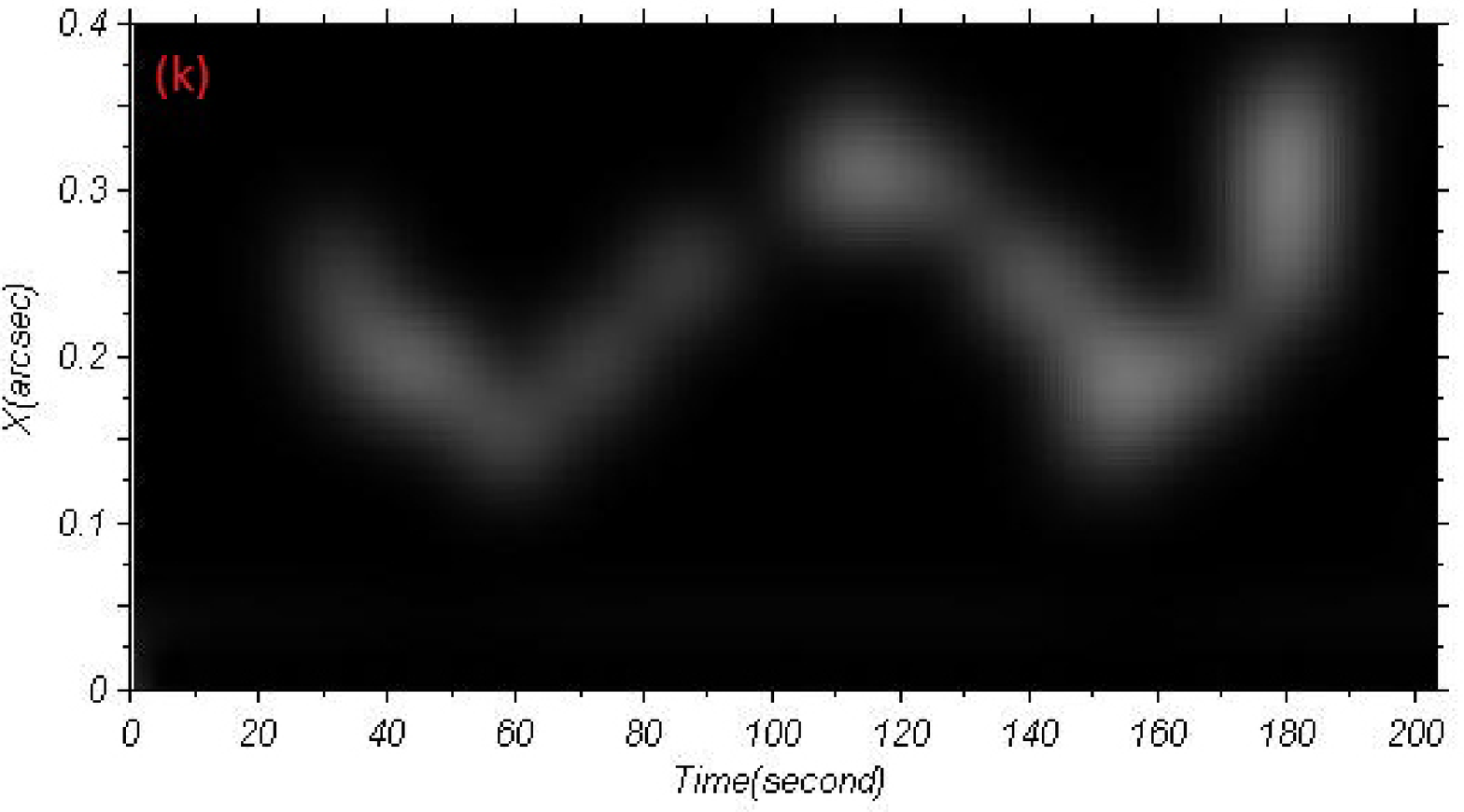}
\includegraphics[height=3cm,width=5cm]{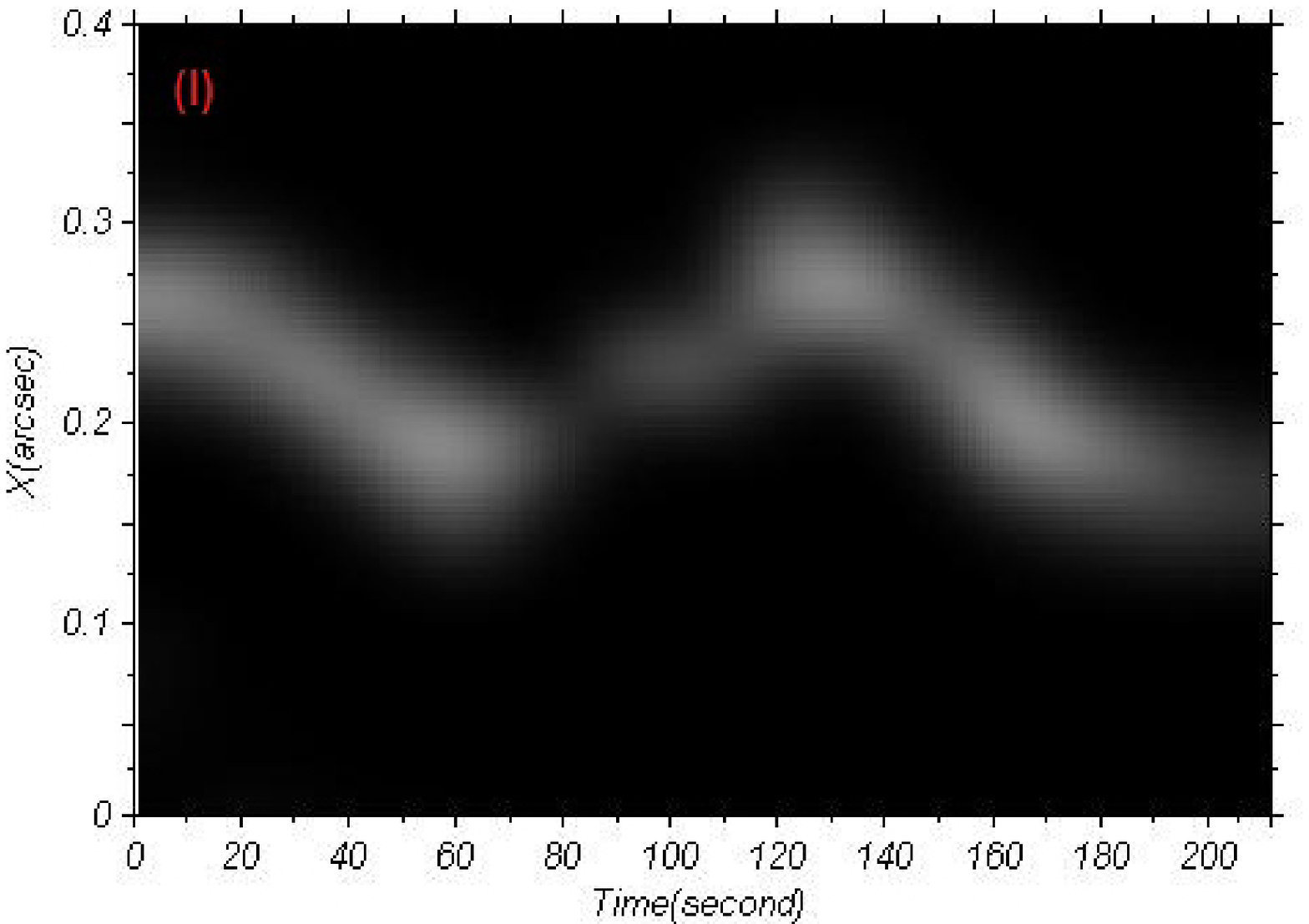}
\includegraphics[height=3cm,width=5cm]{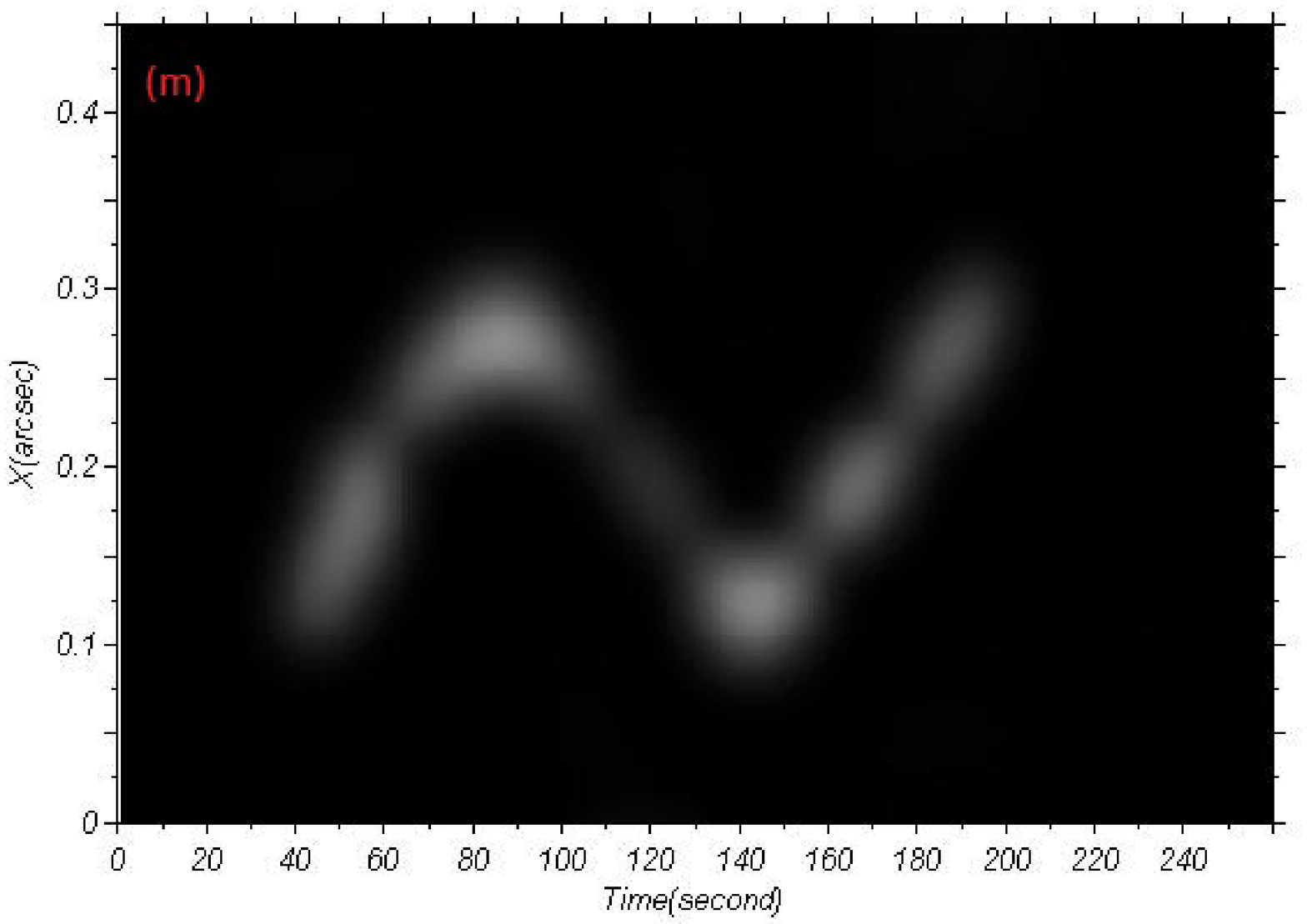}
\includegraphics[height=3cm,width=5cm]{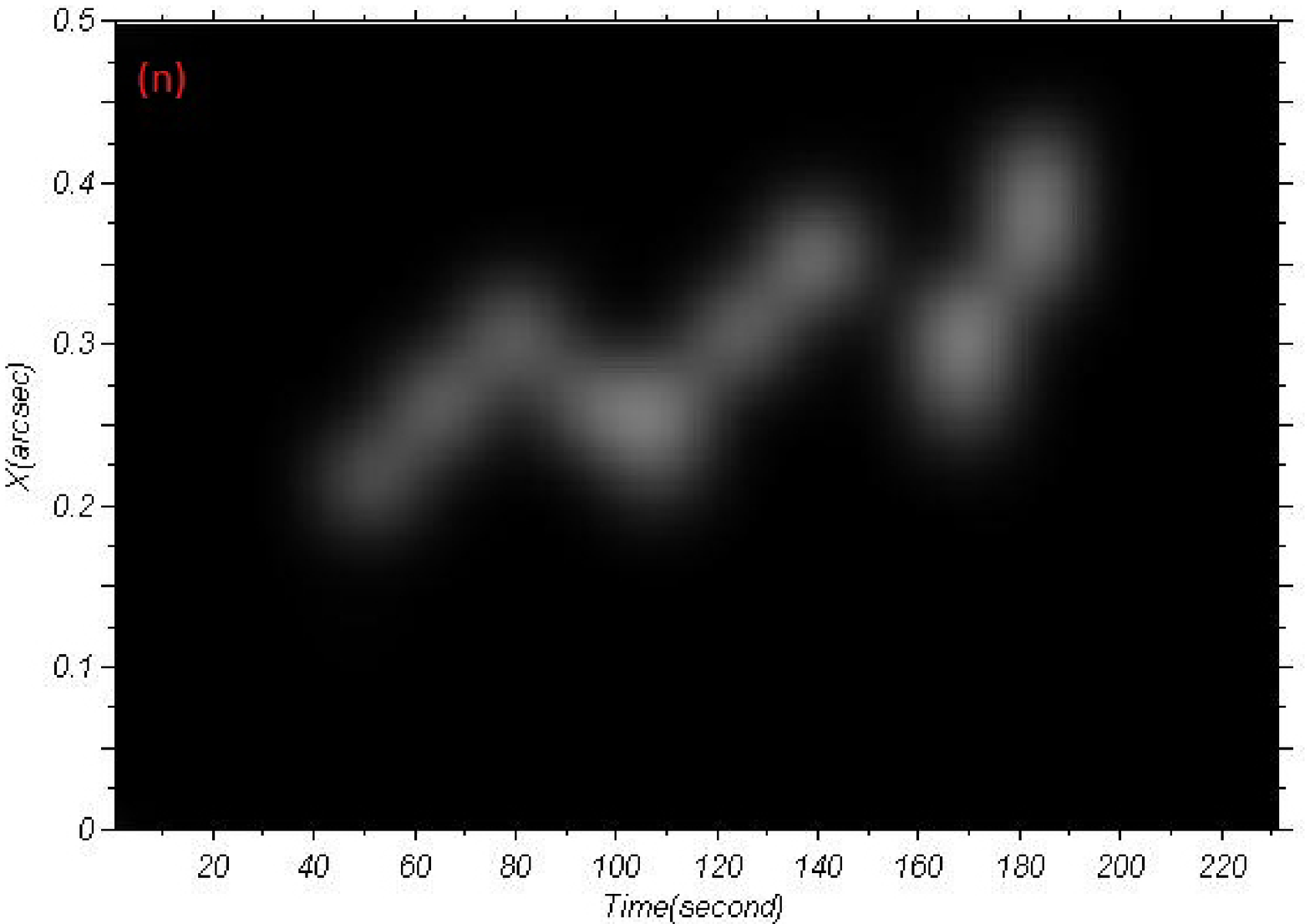}
\caption{ Time slice diagrams of the spicules in \mbox{Ca\,\textsc{ii}} H-line. Axes are in heliocentric coordinates, where 1$''$
\textbf{$\approx$}725km. \label{fig2}}
\end{figure*}

\begin{figure*}
\centering
\includegraphics[height=3.3cm,width=5cm]{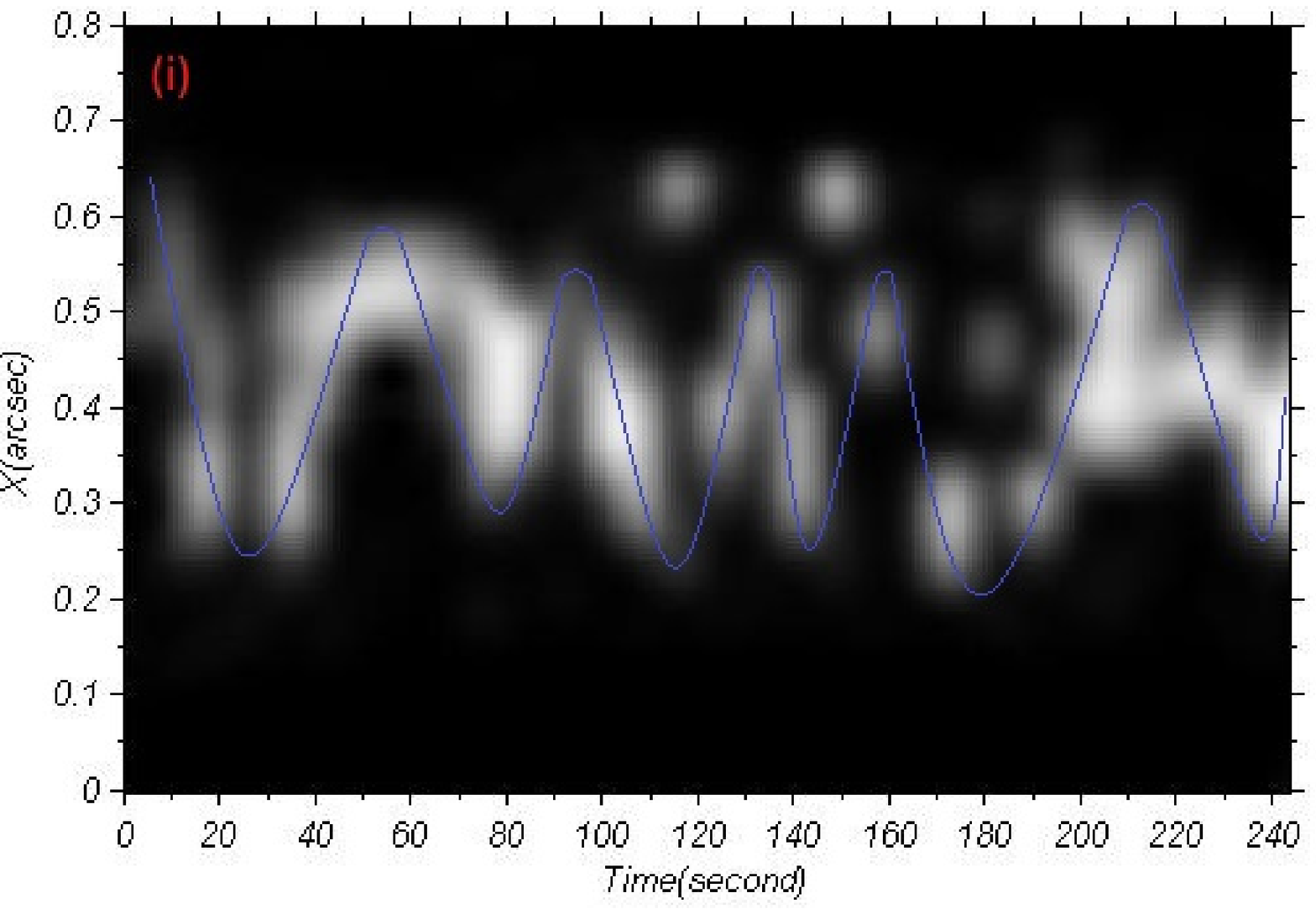}
\includegraphics[height=3.3cm,width=5cm]{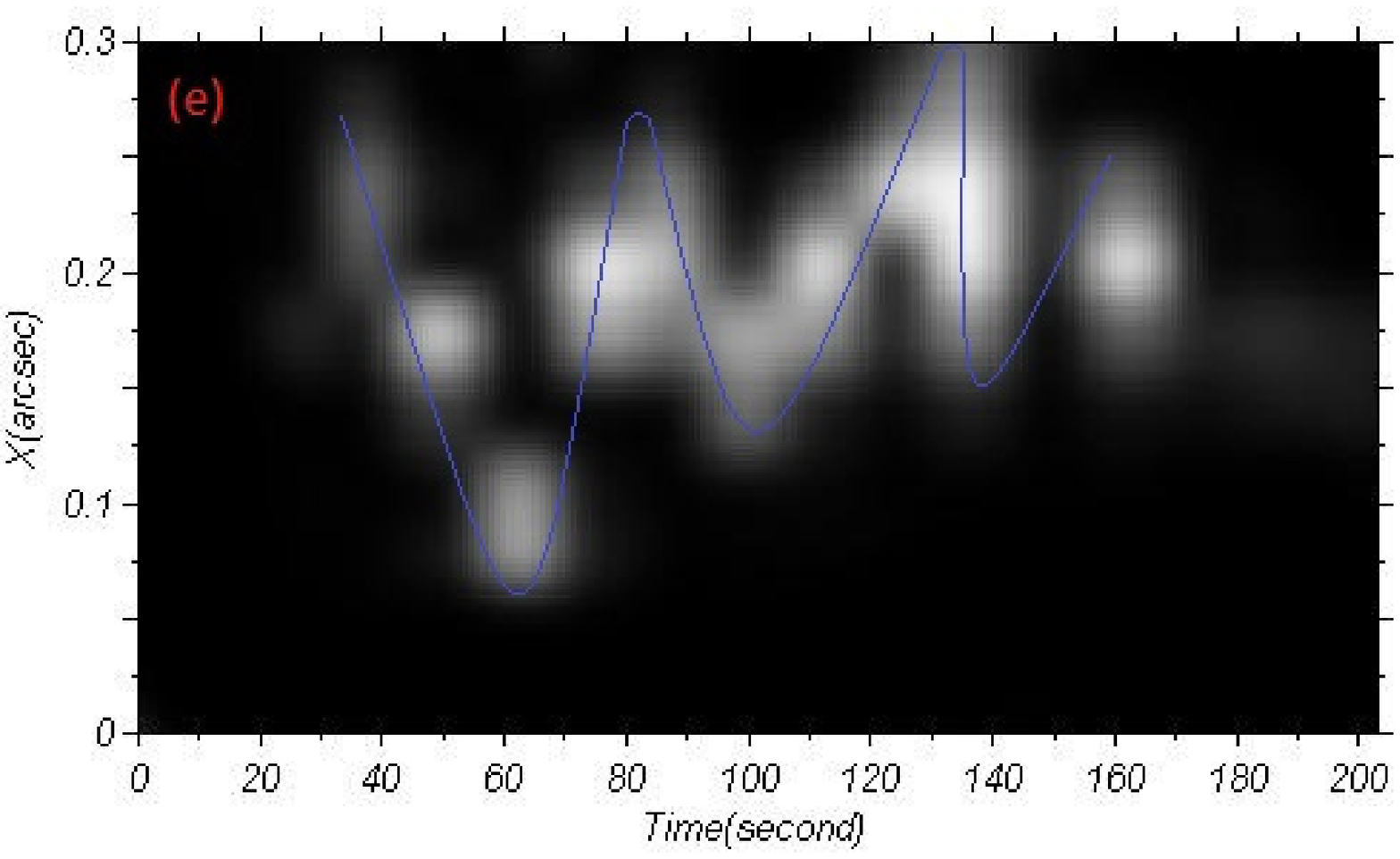}
\includegraphics[height=3.3cm,width=5cm]{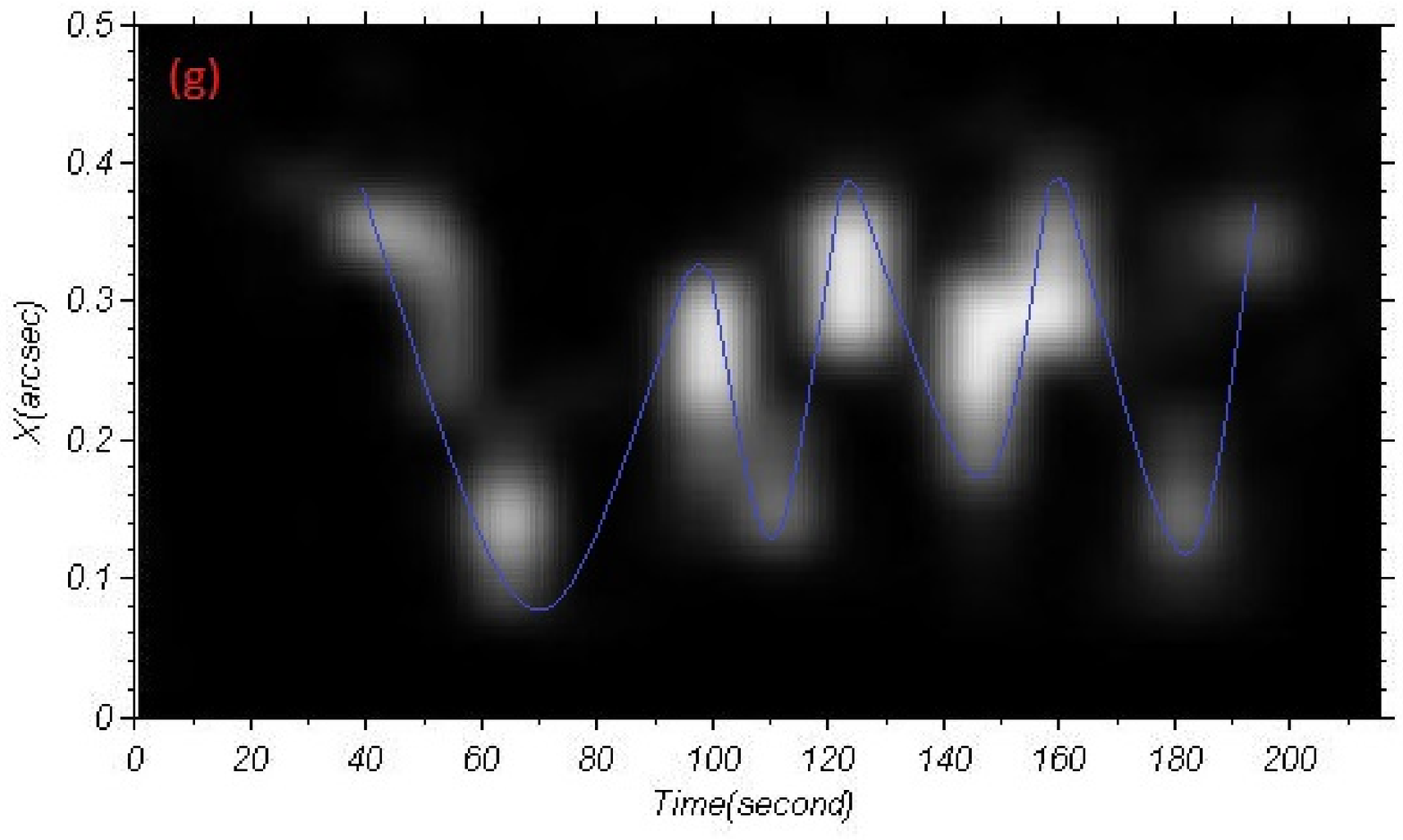}
\caption{ Time slice diagrams of spicule in \mbox{Ca\,\textsc{ii}} H-line. The blue line is a fit to the oscillation of the spicule. \label{fig3}}
\end{figure*}

In Figure~\ref{fig2}, we presented the time slice diagrams of the time series
which consists of 30 consecutive images. The bright regions show the transversal oscillations of spicules axis.
The clear quasi-periodic transverse motion of the spicules
axis is seen on the figure. It should be noted that these diagrams are performed at the same height from the limb.
In Figure~\ref{fig3}, we presented the time slice diagrams and fitted polynomials for three spicules. As it is clear,
the fitted functions are the best ones for our observed results.
We estimated the oscillation period and amplitude at each diagram and presented them in Table 1.
\begin{table*}[htbp]
\centering
\caption{The amplitude ($\xi$), oscillation period($T_{obs}$),  phase speed$(v_{k})$,  magnetic field($B_{0}$) and energy flux(F) for the spicule oscillations.}
   \label{Table1}
\vspace{0.3cm}
\begin{tabular}{|c|| c c c c c|}
\hline
$No$ & $T_{obs}$ (s) & $\xi(arcsec)$ & $v_{k}(kms^{-1})$ & $B_{0}$ (G) & $F(Jm^{-2} s^{-1})$\\
\hline
a & 70  &  0.3 &    57     &     8.9  & 123     \\

b & 70  &  0.2 &    38      &    5.9  & 36      \\

c & 40  &  0.1 &    33      &     5   &  8      \\

d & 50  &  0.1 &    26      &     4   &  7      \\

e & 55  & 0.2  &    48      &   7.5   &   46    \\

f & 90  &  0.4 &    59      &    9.2  &   227   \\

g & 40  & 0.25 &    83      &    13   &  125    \\

h & 100 & 0.35 &    47      &    7.2  &   138   \\

i & 50  &  0.3 &    80      &   12.5  &   172   \\

j & 40  &  0.15&    50      &    7.7  &   27    \\

k & 60  &  0.25&    56      &    8.7  &   83    \\

l & 150 &  0.15&    13      &     2   &    7    \\

m & 100 &  0.2 &    27      &    4.2  &    26   \\

n & 80  &  0.15&    25      &    5.4  &    14   \\
\hline
\end{tabular}
\end{table*}

The density is almost homogeneous along the spicule axis \citep{bek68}; therefore, the density scale height
should be much longer than the spicule length. The variation of oscillation amplitude with height cannot be
due to the decreased density. So, we may have that the oscillation amplitude is proportional to sin(kz),
where k is the oscillation wave number. This expression can be approximated in the long wavelength limit as \textbf{$\sim$} kz.
Hence, the oscillation wavelength is $\lambda$ \textbf{$\sim$} 1/0.06 arc\,sec \textbf{$\sim$} 11500 km \citep{Ebadi2012}.
This allows the estimation of the kink speed as $v_{k} {\sim}$ $13$--$83$ km s$^{-1}$.

It is possible to estimate the Alfv\'{e}n speed and consequently magnetic field strength through them.
Kink waves are transverse oscillations of magnetic tubes and the phase speed for a straight homogenous
tube can be written as:
\begin{equation}
\centering
\label{eq:phasespeed}
v_{k} = \frac{\lambda}{T_{obs}} = v_{A}\sqrt{\frac{\rho_{0}}{\rho_{0}+\rho_{e}}},
\end{equation}

where $v_{A}\equiv B_{0}/\sqrt{\mu_{0}\rho_{0}}$ is the Alfv\'{e}n speed inside the tube, $\lambda$ is the wavelength and $T_{obs}$ is the observed oscillation period.
$\rho_{e}$ and $\rho_{0}$ are the plasma density outside and inside of tube, respectively.
The density in spicules is $3 \times 10^{-10}$ kg m$^{-3}$ and $\rho_{e}/\rho_{0}\simeq 0.02$ \citep{Zaqarashvili2009, Ebadi2012}.

The oscillation amplitude and phase speed allows to estimate
the energy flux, F, storied in the oscillations:

\begin{equation}
\centering
\label{eq:phasespeed}
F = \frac{1}{2}\rho\upsilon^{2} v_{k}
\end{equation}

where $\rho$, $\upsilon$ and $v_{k}$ are density, wave amplitude and phase speed, respectively \citep{Vranjes2008}. The wave velocity can be determined as $v = \xi /T_{obs}$, where $\xi$ is the axis displacement estimated as \textbf{$\sim$} $0.1-0.5$ arcsec and $T_{obs}$ is the oscillation period estimated as \textbf{$\sim$} $40-150$ s.
With these parameters the energy flux is estimated as $F$ = $7$ -- $227$ J\,m$^{-2}$\,s$^{-1}$.

Using the kink speed, the magnetic field strength in spicules at the height of $6000$ km is estimated as $B_{0}$ = $2$ -- $12.5$ G.
This is in good agreement with recently estimated magnetic field strengths in quiet-Sun spicules (\textbf{$\sim$}10G), which was
obtained by spectropolarimetric observations of solar chromosphere in He I $\lambda$ 10830 \citep{Trujillo2005, Singh2007, Ebadi2012}.

\section{Conclusion}
\label{sec:concl}
We performed the analysis of Ca \begin{footnotesize}II\end{footnotesize} H-line time series at the solar limb obtained from \emph{Hinode}/SOT in order to uncover the oscillations in the solar spicules.  We studied 14 random spicules and found that their axis undergo quasi-periodic transverse displacements. The period of the transverse displacements are \textbf{$\sim$} $40-150$ s and the amplitude are \textbf{$\sim$} $0.1-0.5$ arc\,sec. The same periodicity was found in Doppler shift oscillation by \citet{Zaqarashvili2009} and \citet{De2007}, so the periodicity is probably common for spicules. The magnetic field strength in spicules at the height of $6000$ km is estimated as $B_{0}$ = $2$ -- $12.5$ G. These magnetic field strengths are smaller than the values calculated by \citet{Zaqarashvili2009}.

We think that the observed quasi-periodic displacement of spicule axis can be caused due to standing kink waves.
The energy flux storied in the oscillations is estimated as $7$ -- $227$ J\,m$^{-2}$\,s$^{-1}$,
which is of the order of coronal energy losses in quiet Sun regions.

\acknowledgments
The authors are grateful to the Hinode Team for providing the observational
data. Hinode is a Japanese mission developed and lunched
by ISAS/JAXA, with NAOJ as domestic partner and NASA and
STFC(UK) as international partners. Image processing Mad-Max program
was provided by Prof. O. Koutchmy. This work has been supported financially by the Research Institute for Astronomy and
Astrophysics of Maragha (RIAAM), Maragha, Iran.

\makeatletter
\let\clear@thebibliography@page=\relax
\makeatother

\end{document}